\documentclass{article}
\usepackage[T1]{fontenc} 
\usepackage[utf8]{inputenc} 
\usepackage{ismir,amsmath,cite,url}
\usepackage{graphicx}
\usepackage{color}

\usepackage{lineno}
\usepackage{algorithm}
\usepackage{algpseudocode}

\title{MIDI-to-Tab: Guitar Tablature Inference via\\ Masked Language Modeling}

\multauthor
{Drew Edwards \hspace{1cm} Xavier Riley \hspace{1cm} Pedro Sarmento \hspace{1cm} Simon Dixon} 
{Centre for Digital Music,
Queen Mary University of London, UK\\
{\tt\small \{a.c.edwards, j.x.riley, p.p.sarmento, s.e.dixon\}@qmul.ac.uk}
}

\def\authorname{D. Edwards, X. Riley, P. Sarmento, and S. Dixon}

\usepackage[bookmarks=false,pdfauthor={\authorname},pdfsubject={\papersubject},hidelinks]{hyperref}

\begin{document}

\maketitle
\begin{abstract}
Guitar tablatures enrich the structure of traditional music notation by assigning each note to a string and fret of a guitar in a particular tuning, indicating precisely where to play the note on the instrument. The problem of generating tablature from a symbolic music representation involves inferring this string and fret assignment per note across an entire composition or performance. On the guitar, multiple string-fret assignments are possible for most pitches, which leads to a large combinatorial space that prevents exhaustive search approaches. Most modern methods use constraint-based dynamic programming to minimize some cost function (e.g.\ hand position movement). In this work, we introduce a novel deep learning solution to symbolic guitar tablature estimation. We train an encoder-decoder Transformer model in a masked language modeling paradigm to assign notes to strings. The model is first pre-trained on DadaGP, a dataset of over 25K tablatures, and then fine-tuned on a curated set of professionally transcribed guitar performances. Given the subjective nature of assessing tablature quality, we conduct a user study amongst guitarists, wherein we ask participants to rate the playability of multiple versions of tablature for the same four-bar excerpt. The results indicate our system significantly outperforms competing algorithms. 
\end{abstract}

\begin{figure*}[ht]
    \centering
    \includegraphics[width=0.85\textwidth]{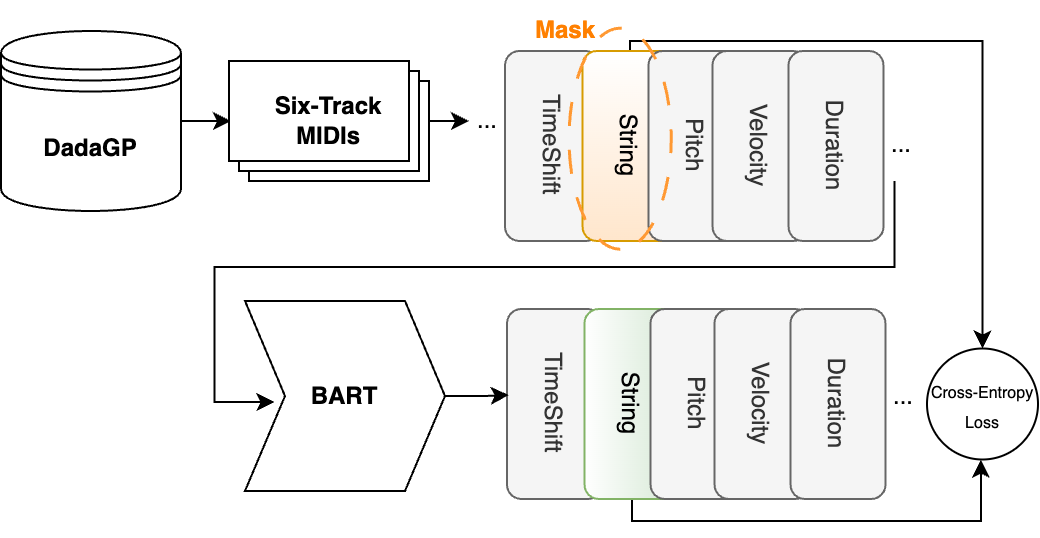}
    \caption{Overview of the training procedure. Guitar Pro files from DadaGP are converted to six-track MIDI files, one file per distinct guitar part and one track per string. These are tokenized into the Structured tokenization of MidiTok. We train a BART model in a simple masked language modeling task where the string tokens are masked out. Only the predictions for the string tokens are used for loss signal propagation.}
    \label{fig:training}
\end{figure*}

\section{Introduction}\label{sec:introduction}

Tablatures (tabs) are a type of music notation where each played note is indicated by its physical position on the instrument, as opposed to merely its pitch. Whereas on (e.g.) the piano, each pitch can be played in exactly one location on the instrument, most playable pitches on stringed instruments like the guitar or violin can be played in multiple positions \cite{Sarmento2023ShredGP}. This redundancy introduces an additional layer of analysis to derive mechanics of a performance from raw pitches. In traditional music scores, e.g.\ for classical guitar, it is the burden of the performer to select appropriate fingerings and positions for the notes in the sheet music. Similarly, a MIDI transcription of a guitar recording lacks this crucial information for a guitarist to replicate the performance. 

In this research we examine the problem of mapping a symbolic representation of a musical performance to guitar tablature. There are few recent publications on this topic (see Section \ref{sec:related-work}), although there are commercial solutions available. Most existing methods propose a manually defined objective function, often related to the difficulty of hand stretches to play chords and distances between hand positions, and seek a solution that minimizes the cost. We take a different approach and provide a modern machine learning treatment of the problem.

We cover the following aspects of our research in this paper: first, we provide a background on the research related to guitar transcription and tablature estimation. Then we formally define the problem of tablature inference from symbolic music notation. Next, we describe the methods of our research, which include: a simple tokenization, a masked language model learning task, a Transformer model solution, pre-training and fine-tuning phases, and a custom beam search inference. Finally, we characterize the performance of our system with quantitative and qualitative metrics, including a detailed user study with 15 guitarists rating various tablatures for short solo guitar excerpts. Our results indicate that guitarists significantly prefer our automatic tablatures versus the commercial alternatives we benchmark. 

\section{Related Work}\label{sec:related-work}

The earliest algorithmic approaches to automatic guitar tablature systems date back to Sayegh \cite{Sayegh1989}. His approaches include an expert system approach assigning rules of permissible transitions between hand positions. These are encoded and enforced via Prolog and its native constraint solver. The second approach described assigns costs to transitions between fingerings and uses dynamic programming (Viterbi \cite{Viterbi}) to find an optimal path through the constructed weighted graph. This latter approach represents the standard classical benchmark for tablature inference. Alternate approaches include genetic algorithms \cite{Genetic}, hybrid expert systems \cite{S2T}, and hidden Markov models \cite{hori2013}. Hori and Sagayama \cite{Hori2016MinimaxVA} extend the dynamic programming approach by finding a path that minimizes the maximum cost of a local transition across a phrase, as opposed to minimizing the global cost as in Sayegh. Radicioni \cite{Rad05} estimates the fingers employed as well as the fret-string combinations, using a graph search paradigm to optimize the bio-mechanical comfort of rendering a piece.

In addition to purely symbolic approaches, there is considerable research on the topic of automatic guitar transcription from audio input. Yazawa et al. \cite{TwoStageTranscription} follow a two-stage approach which uses latent harmonic allocation for multi-pitch estimation (MPE) and then removes unplayable pitches as determined by a fingering cost algorithm similar to Sayegh. Wiggins and Kim \cite{andrew_wiggins_2019_3527800} apply a convolutional neural network to jointly perform MPE and tablature fingering. The strongest performing MPE methods for guitar \cite{tfperceiver, HighResolution} leverage the vast amount of transcription material from other instruments (particularly piano) to enlarge the training dataset, but they fall short in offering no tablature estimations. 

The most similar approach to our own is described in the master's thesis of Mistler \cite{Mistler}, where recurrent neural networks are trained to predict guitar tablatures. However, the training dataset used only contained 74 songs and uses hand-crafted features extracted from the input MusicXML. In contrast, we train on tens of thousands of tabs and process a raw, MIDI-derived tokenization of the input score. The data used for training our network comes from the DadaGP dataset \cite{DadaGP}, comprising 26,181 song scores in the Guitar Pro format.

\section{Problem Formulation}\label{formulation}

We simplify the task of guitar tablature estimation to the task of assigning notes to strings. For a specific guitar tuning, the combination of pitch, string, and fret has only two degrees of freedom. Thus, since the pitch is known a priori, we may predict the string and compute the resulting fret for the assignment. This essentially reduces the problem to \emph{sequence labeling}. In order to increase the flexibility of our system to process a variety of data sources, we begin with MIDI data. Any digital score can be converted to MIDI, and most automatic transcription systems produce MIDI data as well, enabling our tablature system to be composed with any MPE algorithm. 

The problem is formally structured as follows: given a one-track MIDI file $M$, the system produces a six-track MIDI file $M_{S}$, where each track contains the notes assigned to a particular string. Let 

\[ \mathcal{O} = \{64, 59, 55, 50, 45, 40\} \] 

\noindent denote the list of MIDI note numbers for the open strings of the guitar in standard tuning, corresponding to \(E_4, B_3, G_3, D_3, A_2, E_2\), respectively. Thus, to derive the fret of a note with MIDI note number \(n\) assigned to a string \(s\), where \(s \in \{1, 2, 3, 4, 5, 6\}\), the fret \(f\) is calculated by:

\[ f = n - \mathcal{O}[s] \]

Although our derivation of the fret value assumes a standard tuning, this approach could be easily modified to permit alternate tunings by changing the values of $\mathcal{O}$.

\section{Methods}\label{solution}

\begin{figure*}[ht]
    \centering
    \includegraphics[width=0.9\textwidth]{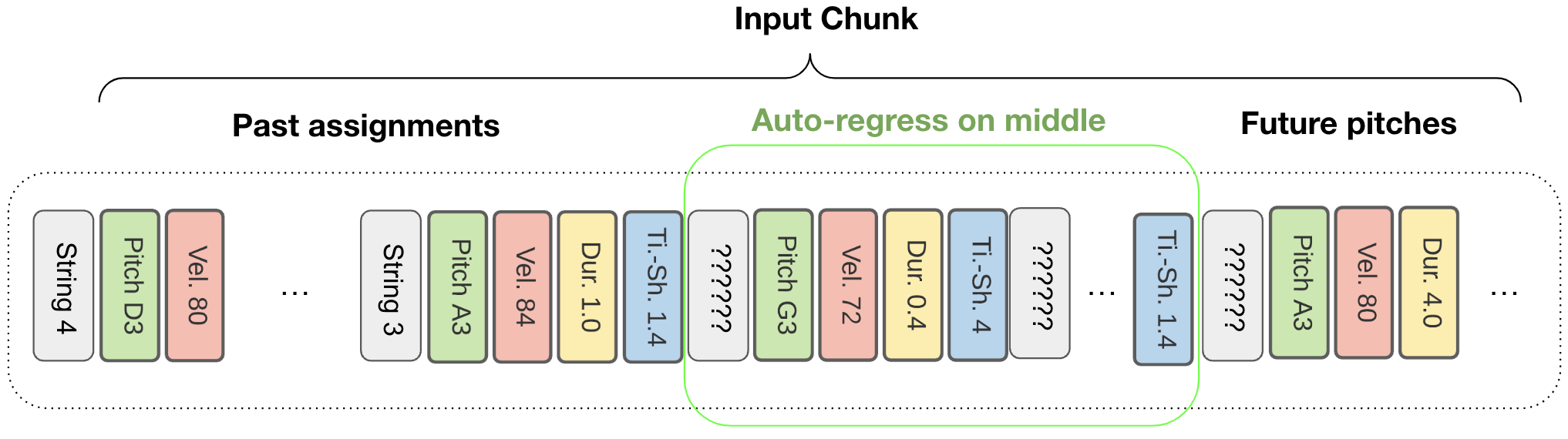}
    \caption{A diagram of our quintile inference algorithm. The middle fifth of the attention window is predicted in an auto-regressive fashion. String assignments from earlier quintiles are fixed. Future notes are available in the context window but will not be assigned until the processing window places them in the center. The beam search is not depicted.}
    \label{fig:quintile}
\end{figure*}
\subsection{Architecture}

Our solution to the problem uses a Transformer with a bi-directional auto-encoder and a left-to-right decoder, based on the BART model architecture \cite{Lewis2019BARTDS}. Using such an approach requires a tokenization of the input data. For this, we use the \emph{Structured} tokenization scheme of Huang and Yang \cite{PopMusicTransformer}. For each MIDI note, we produce five tokens: time shift, string (i.e.\ track), pitch, velocity, and duration. 

During training, we use a simple masked language modeling supervision scheme, masking the string tokens. This permits both past and future note values to be available for the network during inference, but only past string assignments can be seen. During training, we only compute the loss for string token predictions. We use the Hugging Face Transformers package to define the network, using hyperparmeter settings that simply halve the \texttt{bert-base} configuration: 384-dimensional hidden size, 6 hidden layers, 6 attention heads, 1536-dimensional intermediate size, dropout probability of 10\%. These hyperparameter values were not finetuned.

\subsection{Training}
Training takes a two-phase approach (see Figure \ref{fig:training}): first we train from scratch on 27,619 guitar tablatures derived from DadaGP\footnote{Some pieces in DadaGP have multiple guitar parts, and some tracks are filtered out in the conversion process.}. We use the pre-processing code from the SynthTab project \cite{synthtab2024} to produce a six-track MIDI file for each guitar part in the Guitar Pro files. Training is performed with the AdamW optimizer of PyTorch, employing a linear decay schedule with warm-up and an initial learning rate of $1 \times 10^{-4}$ and runs for 100 epochs. The second phase of training is a finetuning on precisely annotated guitar performances from the training splits of Riley et al. \cite{HighResolution} and GuitarSet \cite{guitarset}. The fine-tuning stage is motivated by the concern of data quality in the DadaGP annotations, which were scraped from the online, crowd-sourced tab library Ultimate Guitar\footnote{\url{https://www.ultimate-guitar.com/}}. Here we fine-tune with a learning rate of $1 \times 10^{-5}$, again for 100 epochs, on the much smaller data of 281 tabs. Examples are fed into the network in note sequences of length 50, corresponding to 250 tokens per example. 

\subsection{Inference}
A common problem when training Transformer models for auto-regressive tasks is an asymmetry between training and inference regarding previously predicted sequence values. To leverage the parallelism of the Transformer architecture, ground truth labels must be used as decoder input for masked preceding values. However, during inference on unseen data, these labels are unavailable.

We implement a novel inference mechanism for our algorithm. We break up the input segment into quintiles (10 notes or 50 tokens per quintile). Excluding boundary cases, we only make predictions for the center quintile (see Figure \ref{fig:quintile}). This allows our network to have the ability to see the 20 previous note-string assignments and the next 20 future note values. The attention window is advanced by 10 notes per inference step. Additionally, we implement a custom beam search inference. For each string prediction in a quintile, we retain the top two string values for the note. We limit the number of potential paths to 32. The paths are batched to keep inference times nearly equivalent to naive autoregression. Paths are pruned by taking the maximum probability computed by summing the logits of the string predictions. While this does not fully resolve the asymmetry between inference and training, the beam search and additional context provide more probable decoder input values than naive autoregression.

\subsection{Post-processing}
Thus far we have not imposed any constraints on the output of the network. Ideally, we would take the string predictions and directly augment the score information with the resulting tablature. However, in our qualitative assessment of the system, there are occasions where a string-fret prediction can lead to invalid or unplayable notes. To address these outliers, we attempt to relocate the note to a more suitable string. 

\begin{figure*}[ht]
    \includegraphics[width=1.0\textwidth]{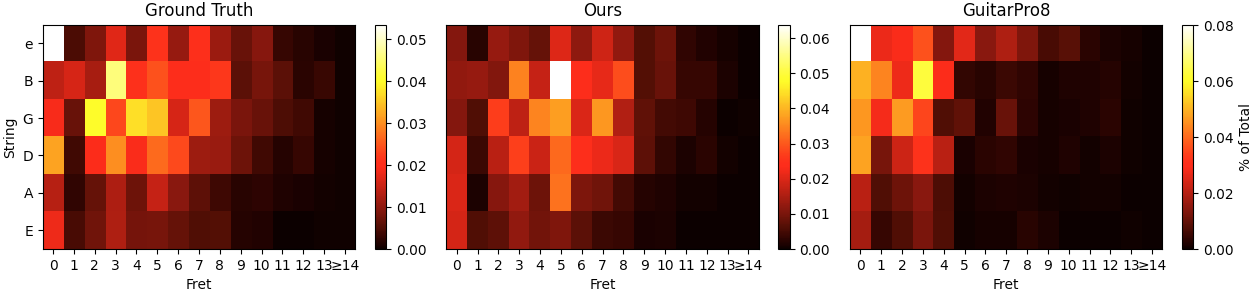}
    \caption{Heatmaps of the fret-string distributions for three of the five tablature systems (ground truth, ours, and Guitar Pro 8). Overall, our system has a similar distribution to the ground truth, but the output appears to be biased away from open strings. Guitar Pro 8 shows a heavy skew to low frets, which perhaps suggest a bias towards playing in ``first position'' (playing primarily on frets 1 to 4).}
    \label{fig:heatmaps}
\end{figure*}

The heuristic algorithm is as follows:
\begin{enumerate}
    \item Merge and sort notes from all strings by start time.
    \item Set maximum allowable deviation from the average fret position (\textit{MAX\_DEVIATION} = 5) and the highest playable fret (\textit{MAX\_FRET} = 21).
    \item For each run of 11 notes (5 past, 1 middle, 5 future):
    \begin{enumerate}
        \item Find the average fret value of the run, excluding open strings from the computation.
        \item If the middle note has a fret value exceeding \textit{MAX\_FRET} or  \textit{MAX\_DEVIATION}:
        \begin{enumerate}
            \item Define the available strings to be those with no notes intersecting with the note under consideration.
            \item If an open string is available, select it.
            \item Else, select the string yielding a fret value closest to the neighborhood mean.
        \end{enumerate}
    \end{enumerate}
\end{enumerate}

\noindent Ultimately, in our test set from Riley et al. \cite{HighResolution}, our post-processing algorithm only modifies 0.53\% of all notes. However rare, addressing these failures is important to ensure the resulting tab is playable.

\section{Results}\label{results}
\subsection{Quantitative Results}
We use the training splits from  Riley et al. \cite{HighResolution}, with 61 pieces in the training set, 8 in the validation set, and 9 in the test set, corresponding to 58,080, 7,031, and 8,451 notes respectively. At the end of finetuning, our next note accuracy on the validation set is 94.35\%. This is the probability of correctly inferring the next note-string assignment given ground truth labels up to the point of prediction. When evaluating autoregressively on the held-out test set across 50-note\footnote{Recall the model has a context window of 50 notes.} examples, our network agrees with ground truth on 82.52\% of predictions. This discrepancy highlights the difference between teacher-forcing and errors accumulated in auto-regressive inference. We measure the impact of the finetuning step by evaluating the pre-trained model without finetuning, which gives 78.48\% agreement, corresponding to a 4.04 percentage point difference due to finetuning.

\begin{table}[b]
\caption{Summary of quantitative analysis, showing maximum, mean and median ``stretch'' of chords in the tablature, defined as the maximum fret distance between any two notes in the chord. We also report the percent agreement with the ground truth note-string assignment. All metrics are averaged over the entire withheld test set.}
\vspace*{1mm}
\label{tab:chord_stretch_stats}
\begin{tabular}{lcccc}
\hline
\textbf{Source} & \textbf{Max} & \textbf{Mean} & \textbf{Median} & \textbf{\%} \\
 & \textbf{Stretch} & \textbf{Stretch} & \textbf{Stretch} & \textbf{Agree} \\
\hline
Ground Truth & 6 & 1.04 & 0 & -- \\
Ours & 12 & 1.84 & 1 & 73.58 \\
MuseScore & 10 & 1.19 & 1 & 62.51 \\
Guitar Pro 8 & 12 & 0.78 & 0 & 62.27 \\
TuxGuitar & 18 & 2.03 & 1 & 55.42 \\
\hline
\end{tabular}
\end{table}

To compare our algorithm to existing technologies, we use one commercially available and two open-source implementations of automatic tablature systems. Guitar Pro 8\footnote{\url{https://www.guitar-pro.com/}} is a music software program designed for editing, visualizing, and sharing guitar, bass, and other stringed instruments' tablatures, and includes an algorithm to automatically produce tablature from score or MIDI. MuseScore\footnote{\url{https://musescore.com/}} is an open-source score editor with a similar functionality for generating tablature. TuxGuitar\footnote{\url{https://www.tuxguitar.app/}} is free, open-source software for creating and playing guitar tablature and standard musical notation. We use each of these systems to generate MusicXML files with tablature for our 9 held-out test scores from our finetuning dataset, which are then used for evaluation.

Objective evaluation of guitar tablature is difficult, as we will discuss further in Section \ref{userstudy}. We provide three metrics that illustrate the strength of our system. The first metric is a measure of agreement between the ground truth note-string assignment and each algorithm's assignment for the corresponding note. The metric is computed by matching\footnote{Matching is required due to reordering of simultaneous notes.} notes from each measure of the ground truth with the notes from the inferred tablature's corresponding measure. An agreement occurs when the ground truth and the inferred tab assign the same string. The total number of agreements is counted across all examples and then divided by the total number of notes compared. Our system shows the highest agreement of 73.18\% (see Table \ref{tab:chord_stretch_stats}). This falls below our 84.42\% agreement from the 50-note examples, because early disagreements on fretboard location for a group of notes will likely cause subsequent note assignments to continue to disagree.

\begin{figure}
    \includegraphics[width=0.45\textwidth]{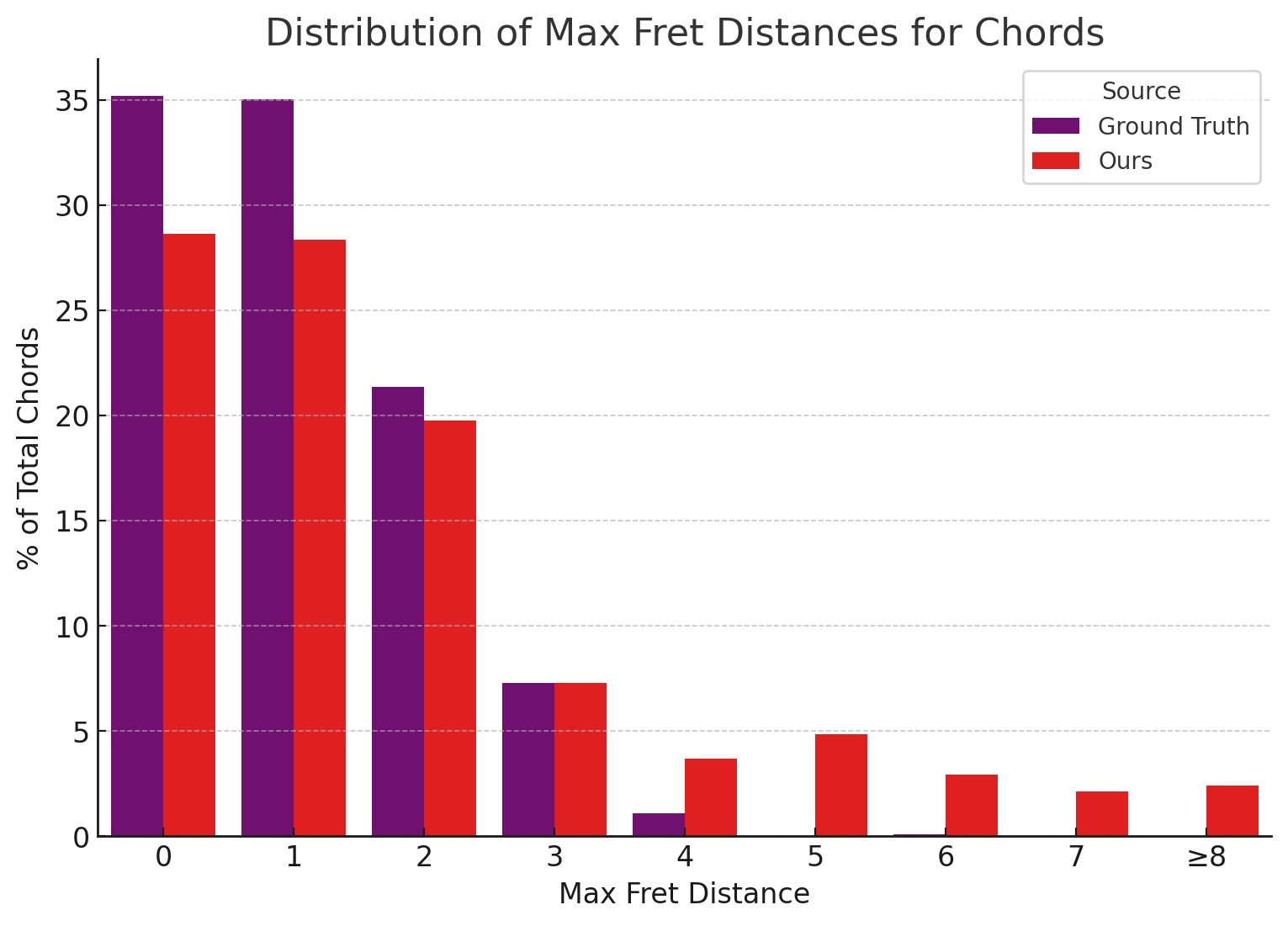}
    \caption{Comparison of the distributions of stretch distances between chords in the test set.}
    \label{fig:stretch-dist}
\end{figure}

\begin{figure}
    \centering
    \includegraphics[width=0.4\textwidth]{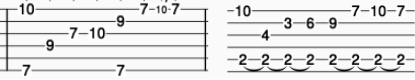}
    \caption{An example failure of our system. Ground truth is left, ours is right. The assignment of B2 to the fifth string creates an 8-fret stretch, which is essentially unplayable.}
    \label{fig:big-stretch}
\end{figure}

The other two metrics relate to the ``\emph{stretch}'' values across chords in the MusicXML. Chords are extracted as note onsets occurring at the exact same time. For all such groups of notes, we define the stretch as the maximum fretwise distance between any two notes in the chord. For example, a chord with notes F3, C4, E4, and A4 played on the string-fret locations\footnote{String 1 is the High E string, and fret values start at 0 for open strings.} (4, 3), (3, 5), (2, 5), (1, 5) will have a stretch value of 2. Open strings do not restrict hand positions so they do not contribute to the stretch. We report the maximum and average stretch across the chords in the test set. The ground truth has the lowest maximum stretch of 6. Our system demonstrates occasional erratic behavior of assigning high notes to lower strings, resulting in a shifted mean and larger maximum stretch value of 12. In Figure \ref{fig:stretch-dist}, we compare frequencies of maximum fret distances between our system and the ground truth. An example failure is shown in Figure \ref{fig:big-stretch}. The mean and median values indicate that, in general, all algorithms attempt to place chords within a narrow band of frets. The presence of these large stretches motivates future work to better inform our algorithm about the importance of physical playability. 

As a final quantitative comparison, we compute fret-string distributions for all five sets of tablatures. For each distribution, we compute the Kullback–Leibler divergence against the ground truth distribution. Our system has the lowest value of 0.099; the other values are: 0.462 for Guitar Pro 8, 0.635 for MuseScore, and 1.286 for TuxGuitar. Three of these distributions are shown in Figure \ref{fig:heatmaps}.

\subsection{User Study} \label{userstudy}

\begin{figure}[b]
    \centering
    \includegraphics[width=0.45\textwidth]{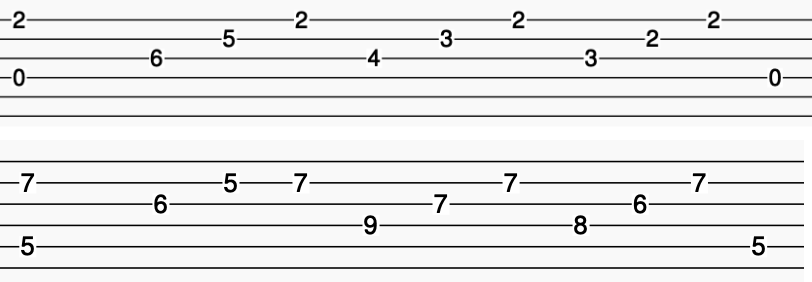}
    \caption{Two tablatures for the same musical excerpt. The top is ground truth, the bottom is from our system. }
    \label{fig:tab_example}
\end{figure}

A purely quantitative evaluation of automatic guitar tablature systems is problematic because there may be multiple ways to play the same phrase or excerpt of music. For example, in Figure \ref{fig:tab_example}, we show two distinct tablatures for the same one-bar phrase. The top is the ground truth transcription and the bottom is the layout from our system. At a glance, both provide reasonable fretboard fingerings. The ground truth shows a preference for open strings, but our system better minimizes the maximum span between successive notes (2 frets versus 4 frets). However, in this example, our system only agrees with the ground truth on two notes, which corresponds to an accuracy of 16.67\%. On the other hand, hand-crafted metrics (such as maximum or average span between notes) fail to capture complex preferences of guitar tablature -- otherwise existing systems that minimize these values as cost functions would suffice. 

To complement the quantitative analysis and circumvent some of the potential limitations of the approach, we conducted a study to assess guitarists' opinions on the \textit{playability} and overall preferences for tablatures. Participants were exposed to 30 audio excerpts consisting of 4-bars of solo jazz guitar audio, and for each were shown 5 distinct tablature transcriptions from the following groups: TuxGuitar (\textit{TG}), MuseScore (\textit{MS}), Guitar Pro 8 (\textit{GP}), our system (\textit{Ours}) and ground truth (\textit{GT}), which was created by a professional transcriber. The stimuli were selected by randomly sampling the test split of Riley et al. \cite{HighResolution}. Via an online listening study, we probed how guitar players deem the tablatures generated by our system, and how they rank them against the ground truth and the outputs from other tablature generation software (i.e.\ \textit{TG}, \textit{MS} and \textit{GP}). 

\begin{figure}[ht]
    \centering
    \includegraphics[width=0.45\textwidth]{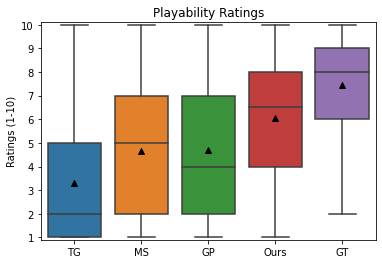}
    \caption{Box plots of the results of the listening study on the playability of tablatures. Bars indicate median values and triangles indicate mean values, for each group. }
    \label{fig:box-plots}
\end{figure}

The online listening study took approximately 1.5 hours to complete and both the order of audio excerpts and the order of tablature transcriptions were randomized. As conditions to take part in the study we proposed that participants should be guitar players and have a familiarity with reading tablatures, access to headphones or speakers and normal hearing. Subjects were instructed to ignore the difficulty of the music excerpts as they rated the playability of the tablatures. Overall, we recruited 15 guitarists and invited them to attempt to play each of the tablature examples on the guitar during the study, while rating each of the tablature transcription groups on a scale from 1 to 10. Participants, with an age distribution of $39 \pm 14$ years, reported a median value of 10 years of daily regular engagement with practice of the guitar, and a median value of 3 hours of guitar practice per day at the peak of their interest. The study received ethical approval from the Queen Mary University of London Ethics of Research Committee (QMERC20.565.DSEECS24.012), and participants were compensated with an 
Amazon gift voucher. Results for the listening test can be observed in Figure \ref{fig:box-plots}.

As expected, the ground truth group ranks highest ($7.45 \pm 2.62$). We hypothesise that the reason why the ground truth falls well short of a ``perfect'' score is linked to the subjective preferences of participants in terms of fingerings and note position choices, which are inherently linked to their guitar playing techniques and overall style. Furthermore, the ground truth represents professional transcriptions of jazz recordings, where perfect information of the original tablature is not available. As discussed above and illustrated in Figure \ref{fig:tab_example}, there can be multiple reasonable ways to play a phrase, and we observed that the ratings for our system were higher than those of the ground truth group for 101 ratings out of 450 in total. The results show that participants tend to rate the playability and overall preference of the tablatures from our system ($6.04 \pm 2.67$) higher than the ones from the competing software (\textit{TG}: $3.32 \pm 2.87$; \textit{MS}: $4.67 \pm 2.88$; \textit{GP}: $4.69 \pm 2.64$). 

Of the 2,250 data points collected ($30$ excerpts  $\times$ $5$ tabs $\times$ $15$ participants), a Shapiro-Wilk test showed that data was not normally distributed within groups. Due to the repeated measurements characteristic of the test (every participant rates all the stimuli), we use a Friedman test to investigate the effects of the type of tablature transcription system on the perceived playability of tablatures, with a Type I error $\alpha$ of 0.05. The statistical results showed a highly significant effect of the tablature transcription system in the participants' responses amongst groups ($\chi^2(4)=532.09$, $p<.001$). Finally, in order to determine if there were statistically significant differences between groups, we conducted a post-hoc pairwise Wilcoxon test, Bonferroni-adjusted $\alpha$ level of $0.005$ ($.05/10$). This yielded highly significant differences in ratings between groups, except for (\textit{MS}, \textit{GP}).

\section{Discussion}

Our results suggest a data-driven approach to guitar tablature inference can yield predictions that are significantly more aligned with guitarists' preferences than existing methods. These results are very encouraging given the simplicity of our approach. Our system imposes no constraints on the predicted tablatures until the final post-processing, during which less than 1\% of note-string assignments are modified. Future research in this direction may benefit from more directly encoding positional fretboard locations and physical limitations as input to the network. 

Despite the strong results, our system has several limitations to be addressed. Guitar tuning is never explicitly encoded as input to the model. Since we only predict string values, our fret predictions are always derived from the note-string assignment and an assumption of standard tuning. Similarly, our system does not handle the use of capos. The system is unaware of many guitar specific articulations, such as harmonics, hammer-ons, pull-offs, and pitch bends. Finally, we make no attempt to assign individual notes to the fingers of a guitarist, which is occasionally done in professional scores or transcriptions, and would be a necessary step in order to estimate playability explicitly.

Another criticism of our approach is that it does not use visual and audio cues for fretboard prediction. As shown by Bastas et al.\ \cite{Bastas}, inharmonicity analysis of a particular instrument can improve string predictions. Likewise, Duke and Salgian \cite{cvtabs} demonstrate how computer vision models can be used for accurate and real-time tablature transcription. Both of these directions of research offer a more faithful reproduction of a particular performance, since a symbolic approach simply has no access to disambiguating signals such as hand position or string inharmonicity. However, this shortcoming can also be viewed as a strength: our system does not need access to video nor audio. From this perspective, our approach can be viewed as an automatic arranging system for guitar tablature performance.

The main failure mode of our system is the assignment of unplayable chords at a small but significant frequency (2.4\% of chords have a maximum fret distance exceeding 7). Future research may explore different tokenization schemes: encoding fret values as input, physically inspired loss functions, or more carefully designed post-processing to handle these cases. However, the vast majority of the mass of the distribution of chord stretch distances falls within playable limits, which indicates that the algorithm is implicitly modeling some of the physical constraints that classical systems use to derive tablatures.


\section{Conclusion}

We present a deep learning algorithm to predict guitar tablature from symbolic music notation. Our methodology trains an encoder-decoder Transformer to learn tablature assignment from raw note events. Drawing inspiration from natural language processing, we begin by pre-training on a dataset of tens of thousands of tablatures and then fine-tune on a curated dataset of professional guitar scores. We evaluate our system against commercially available software and demonstrate a significant preference for our system through a user study among guitarists. Our MIDI-to-Tab system represents a first step towards achieving human-level tablature inference via machine learning.

\newpage

\section{Acknowledgments}
Authors DE, XR, and PS are research students at the UKRI Centre for Doctoral Training in Artificial Intelligence and Music, supported by UK Research and Innovation [grant number EP/S022694/1] and Yamaha Corporation (DE).

\bibliography{ISMIRtemplate}

%
%
%
%
%

\end{document}